\documentclass[amsfonts,aps,nofootinbib,preprint]{revtex4}

\begin{document}

\title{Observer dependent D-brane for strings propagating in pp-wave time dependent background}

\author{D\'afni Z. Marchioro\footnote{dafnimarchioro@unipampa.edu.br} and Daniel
L. Nedel\footnote{dnedel.unipampa@ufpel.edu.br}}

\affiliation{Universidade Federal do Pampa - UNIPAMPA, Rua Carlos Barbosa s/n, Bairro Get\'{u}lio Vargas, CEP 96412-420, Bag\'{e} - RS, Brazil}

\begin{abstract}
We study type IIB superstring in the pp-wave time-dependent background, which has a singularity at $t=0$. We show that this background can provide a toy model to study some ideas related to the stretched horizon paradigm and the complementary principle of black holes. To this end, we construct a unitary Bogoliubov generator which relates the asymptotically flat string Hilbert space, defined at $t =\pm \infty$, to the finite time Hilbert space. For asymptotically flat observers, the closed string vacuum close to the singularity appears as a boundary state which is in fact a D-brane described in the closed string channel. However, observers who go with the string towards to the singularity see the original vacuum.
\end{abstract}

\maketitle

\section{Introduction}

One of the outstanding subjects in string theory is its study
in non trivial backgrounds. Such endeavour may provide answers to
important questions involving applications of string theory in black hole physics and in typical quantum cosmology problems, such as the initial conditions, the cosmological singularities and the pre-big bang scenario.

The central issue to study strings in black holes is the stretched horizon paradigm and the black hole complementary principle \cite{suss1}. According to this, an observer who remains outside the horizon can describe the black hole as a membrane, the stretched horizon. By using the Unruh effect and the Hagedorn behavior, the stretched horizon should be placed at a distance from the event horizon related to the string length scale and it absorbs any matter, energy and information which fall onto it. On the other hand, an observer who falls freely into the black hole sees nothing: no loss of information, no stretched horizon.

As it is well known that in the non-perturbative sector (higher curvature) the string theory is described by extended objects, namely p-branes, it is natural to think that the microscopic structure of the stretched horizon may be related to these objects. However it is a very hard task to check this directly in string theory, mainly because the worldsheet model is very non linear when string propagates in curved backgrounds. On the other hand, the p-branes have a flat space description (small string coupling) in terms of hyperplanes where open strings are attached, the D-branes. 

The main goal of this article is to explore the complementary principle ideas in cosmological backgrounds using 
 D-branes described in closed string channel. To this end we study the closed string propagating in pp-wave time dependent background \cite{tsey} \footnote{For a historical development of the subject see, for example, \cite{Duval, amati, horo, brooks, sanchez}.}, which has a singularity at the origin of the light-cone time ($t=0$) and shares many characteristics with the
Gasperini and Veneziano's pre-big bang cosmology \cite{gaspvene}. Also it was shown in \cite{Borunda} that singular homogeneous plane waves arise as the Penrose limit of some spacetime singular metrics. In this background the worldsheet model exhibits many interesting
properties of field theory in curved space,
including the choice of the vacuum, particle creation (in
this case string mode creation), entropy production and loss of unitarity \cite{gdn}. Besides, the string coupling (the dilaton) is small even close to the singularity, and the worldsheet model is soluble. 

We are going to show that this scenario can be seen as a toy model to explore some ideas related to the complementary principle directly in string theory. In fact it will be shown that for asymptotically flat observers the vacuum close to singularity will be related to a boundary state which corresponds to a D-brane in closed string channel. On the other hand, observers which pass through the singularity together with the string see the original string vacuum. Also, we show that the vacuum of asymptotically flat observers and the vacuum of observers who go together with the string towards the singularity are related by a SU(1,1) Bogoliubov transformation.  

This article is divided as follows: in the second section, we present the model; the third section is devoted to construct the $SU(1,1)$ Bogoliubov generator; in the fourth section, we show how flat observers see the vacuum at $t=0$ as a D-brane boundary state; and the conclusions are presented in the last section.  

\section{String in the plane-wave time-dependent background}

Let us start by considering the bosonic sector of type IIB Green-Schwarz superstring in the pp-wave time-dependent background (in Brinkmann coordinates) \cite{tsey}

\begin{eqnarray}
ds^2 = 2dudv -\lambda x^2du^2 + dx^idx^i \nonumber \\
\phi = \phi(u) \:, \:\:\: \lambda = \frac{k}{u^2}
\end{eqnarray}

\noindent where $u$, $v$ are the light-cone coordinates, $x^i$ ($i=1\ldots 8$) are the transverse coordinates and $\phi$ is the dilaton. After fixing the reparametrization invariance choosing the light-cone gauge, the bosonic part of the action takes the form

\begin{equation}
S = -\frac{1}{4\pi\alpha^{'}}\int dt \int^{\pi}_{0}d\sigma (\partial^{a}x^i\partial_{a}x^{j}\delta_{ij} + \frac{k}{t^2}(x^i)^2)
\end{equation}

\noindent The solution of the equations of motion in Brinkmann coordinates can
be written as
\begin{eqnarray}
x^i(\sigma , t) &=& x^i_0(t) + \frac{1}{2} \sqrt{2\alpha^{'}}
\sum^{\infty}_{n=1}\frac{1}{\sqrt{n}}
[Z(2nt)(a^{i}_{n}e^{2in\sigma} + \tilde{a}^{i}_{n}e^{-2in\sigma})
\nonumber
\\
&+& Z^{\ast}(2nt)(a^{i\:\dagger}_{n}e^{-2in\sigma} +
\tilde{a}^{i\:\dagger}_{n}e^{2in\sigma})]
\end{eqnarray}
with
\begin{eqnarray}
Z(2nt) \equiv e^{-i\frac{\pi}{2}\nu} \sqrt{\pi n
t}[J_{\nu-\frac{1}{2}}(2nt) -iY_{\nu-\frac{1}{2}}(2nt)] \:, 
\label{Zdef}
\\
\nu = \frac{1}{2}\left(1+\sqrt{1-4k}\right)\:, 0<k<\frac{1}{4}\:, \\
x^{i}_{0}(t) = \frac{1}{\sqrt{2\nu -1}}(\tilde{x}^{i}t^{1-\nu}
+ 2\alpha^{'}\tilde{p}^{\: i}t^{\nu})
\\
\tilde{x}^{i} = \sqrt{\frac{\alpha^{'}}{2}}(a^{i}_{0} +
a^{i\dagger}_{0}), \qquad \tilde{p}^{\: i} =
\frac{1}{i\sqrt{2\alpha^{'}}}(a^{i}_{0} - a^{i\dagger}_{0})\:,
\end{eqnarray}
where $J_{\nu-\frac{1}{2}}$ and $Y_{\nu-\frac{1}{2}}$ are the Bessel
functions of the first and second type, respectively, and the
creation/annihilation operators satisfy the usual oscillator algebra
\begin{equation}
[a^{i}_{n},a^{j\:\dagger}_{m}] = [ \tilde{a}^{i}_{n},
\tilde{a}^{j\:\dagger}_{m}] = \delta^{ij}\delta_{nm}, \qquad
[a^{i}_{0},a^{j\:\dagger}_{0}] = \delta^{ij}. \label{poisson}
\end{equation}

\noindent The Hamiltonian for the system under consideration can be written as
\begin{equation}
H = \frac{1}{\alpha^{'}p_{v}}\left({\cal H}_{0}(t) +\frac{1}{2}\sum^{\infty}_{n=1}n
[\Omega_{n}(t)(a^{\dagger}_{n}\cdot a_{n} +
\tilde{a}^{\dagger}_{n}\cdot\tilde{a}_{n}) -
B_{n}(t)a_{n}\cdot\tilde{a}_{n} -
B^{\ast}_{n}(t)a^{\dagger}_{n}\cdot\tilde{a}^{\dagger}_{n}]\right)
\label{h0}
\end{equation}

\noindent the dot denoting the scalar product in the transverse space, and

\begin{eqnarray}
\Omega_{n}(t) = \left(1 + \frac{\nu}{4t^{2}n^{2}}\right)\mid Z\mid^{2} +
\mid W\mid^{2} - \frac{\nu}{2nt}(ZW^{\ast} + Z^{\ast}W),
\nonumber
\\
B_{n}(t) = \left(1 + \frac{\nu}{4t^{2}n^{2}}\right)Z^{2} + W^{2}
-\frac{\nu}{t n}Z W.
\end{eqnarray}
where
\begin{equation}
W(2nt) \equiv e^{-i\frac{\pi}{2}\nu} \sqrt{\pi nt}[J_{\nu +
\frac{1}{2}}(2nt) -iY_{\nu + \frac{1}{2}}(2nt)]. \label{Wdef}
\end{equation}
The term ${\cal H}_{0}$ in (\ref{h0})
is the zero-mode part, which is obtained treating the zero mode as a
harmonic oscillator with time-dependent frequency. As expressed in terms of the modes $a_n$, $\tilde{a}_n$, $a_0$ and $a^{\dagger}_0$, the Hamiltonian is non-diagonal. However, in the limit $t=-\infty$ the Hamiltonian is diagonal and the metric is flat. We can define the asymptotically flat vacuum $|0, p_v \rangle$ as the one annihilated by $a_n$, $\tilde{a}_n$ and the flat Hilbert space is built by cyclic operations of $a_n^{\dagger}$, $\tilde{a}_n^{\dagger}$. This is the Hilbert space related to asymptotically flat observers.

In order to diagonalize the Hamiltonian in a finite time, a new set of time-dependent string modes is defined as

\begin{eqnarray}
{\cal A}^{i}_{n}(t) = f_{n}(t)a^{i}_{n} +
g^{\ast}_{n}(t)\tilde{a}^{i\:\dagger}_{n}, \qquad {\cal
A}^{i\dagger}_{n}(t) = f^{\ast}_{n}(t)a^{i\:\dagger}_{n} +
g_{n}(t)\tilde{a}^{i}_{n}, \nonumber
\\
\tilde{\cal A}^{i}_{n}(t) = f_{n}(t)\tilde{a}^{i}_{n} +
g^{\ast}_{n}(t)a^{i\:\dagger}_{n}, \qquad \tilde{\cal
A}^{i\dagger}_{n}(t) =
f^{\ast}_{n}(t)\tilde{a}^{i\:\dagger}_{n} + g_{n}(t)a^{i}_{n},
\label{osc}
\end{eqnarray}
where
\begin{eqnarray}
f_{n}(t) =
\frac{1}{2}\sqrt{\frac{w_n}{n}}e^{2iw_{n}t}
\left[Z(2nt) + \frac{i}{2w_{n}}\dot{Z}(2nt)\right]\:,
\nonumber
\\
g_{n}(t) = \frac{1}{2}\sqrt{\frac{w_n}{n}}e^{-2iw_{n}t}
\left[Z(2nt) - \frac{i}{2w_{n}}\dot{Z}(2nt)\right],
\label{fg}
\end{eqnarray}
and
\begin{eqnarray}
w_{n}(t) = \sqrt{n^{2} + \frac{k}{4t^{2}}}\:, \nonumber\\
\dot{Z}(2nt)=\partial_{t}Z(2nt) = \frac{\nu}{t}Z(2nt)
- 2nW(2nt)\:,
\label{zdot}
\end{eqnarray}
with $Z(2nt)$ and $W(2nt)$ defined in (\ref{Zdef}) and
(\ref{Wdef}). We can see that this diagonalization process is in fact a canonical transformation - a $SU(1,1)$ Bogoliubov transformation - defined by:

\begin{eqnarray}
\left(
\begin{array}{c}
{\cal A}^{i}_{n}(t) \\
\widetilde{{\cal A}}^{i \dagger}_{n}(t)
\end{array}
\right)
&=&{\mathbb B}_{n}(t)\left(
\begin{array}{c}
a^{i}_{n} \\
\widetilde{a}^{i \dagger }_{n}
\end{array}
\right) ,
\nonumber
\\
\left(
\begin{array}{cc}
{\cal A}^{i \dagger}_{n}(t) & -\widetilde{{\cal A}}^{i}_{n}(t)
\end{array}
\right) &=&\left(
\begin{array}{cc}
a^{i \dagger }_{n} & -\widetilde{a}^{i}_{n}
\end{array}
\right) {\mathbb B}^{-1}_{n}(t), \label{tbti}
\end{eqnarray}
where the $SU(1,1)$ matrix transformation is given by
\begin{eqnarray}
{\mathbb B}_{n}(t)=\left(
\begin{array}{cc}
f_{n}(t) & g^{\ast}_{n}(t) \\
g_{n}(t) & f^{*}_{n}(t)
\end{array}
\right) ,\qquad |f_{n}(t)|^{2}-|g_{n}(t)|^{2}=1. \label{tbm}
\end{eqnarray}

\section{The SU(1,1) Bogoliubov Generator}

We are going to show now that it is possible to find a unitary Bogoliubov generator which generates the canonical transformation above and which can be used to construct a finite time Hilbert space from the asymptotically flat  one. The finite time Hilbert space will be related to observers who go together with the string towards to the singularity. Since a  physical closed string state $|\Phi\rangle$  must obey

\begin{equation}
(N-\widetilde{N})|\Phi\rangle=0,
\end{equation}

\noindent where $N,\widetilde{N}$ are the right and left number operators, the unitary Bogoliubov generator, which generates a new physical closed string Hilbert space from the first one, must satisfy
\begin{eqnarray}
\left[G,\left(N-\widetilde{N}\right)\right]=0 \:, \nonumber\\
G^{\dagger}= G \:. 
\label{g}
\end{eqnarray}
The operators
that satisfy the relations (\ref{g}) and can generate a transformation like (\ref{tbti})  have the 
following form 
\begin{eqnarray}
g_{1_{n}} &=&\theta _{1_{n}}\left( a_{n}\cdot \tilde{a}_{n}+\tilde{
a}_{n}^{\dagger }\cdot a_{n}^{\dagger }\right), \nonumber \\
g_{2_{n}} &=&i\theta _{2_{n}}\left( a_{n}\cdot \tilde{a}_{n}-
\tilde{a}_{n}^{\dagger }\cdot a_{n}^{\dagger }\right) , \nonumber \\
g_{3_{n}} &=&\theta _{3_{n}}\left( a_{n}^{\dagger }\cdot a_{n}+
\tilde{a}_{n}^{\dagger }\cdot \tilde{a}_{n}+ tr\delta ^{i j} \right),
\label{generators}
\end{eqnarray}

\noindent where $\theta_1,\theta_2, \theta_3$ are the transformation parameters, which will be related to the functions $g_{n}(t)$ and $f_{n}(t)$ . It is easy to verify that the generators (\ref{generators}) satisfy the 
$SU\left( 1,1 \right)$ algebra
\begin{equation}
\left[ g_{1_{n}},g_{2_{n}}\right]
=-i\Theta _{123}g_{3_{n}},\quad \left[ g_{2_{n}},g_{3_{n}}\right] =i\Theta _{231}g_{1_{n}},\quad \left[ g_{3_{n}},g_{1_{n}}\right] =i\Theta _{312}g_{2_{n}},
\label{su11}
\end{equation}
where we have defined
\begin{equation}
\Theta _{ijk}\equiv 2\frac{\theta _{i_{n}}\theta _{j_{n}}}{\theta _{k_{n}}}.
\label{thetas}
\end{equation}
\noindent So, as we can see from  (\ref{generators}), the most general string Bogoliubov 
transformation takes the following form
\begin{equation}
G=\sum_{n} G_{n} ,
\label{gentransf}
\end{equation}
where 
\begin{equation}
G_{n} =\lambda _{1_{n}}\tilde{a}_{n}^{\dagger }\cdot
a_{n}^{\dagger }-\lambda _{2_{n}}a_{n}\cdot \tilde{a}_{n}+\lambda
_{3_{n}}\left( a_{n}^{\dagger }\cdot a_{n}+\tilde{a}_{n}^{\dagger }\cdot 
\tilde{a}_{n}+ tr\delta^{ij} \right) ,
\label{rlgen}
\end{equation}
and the coefficients $\lambda_1$, $\lambda_2$ and $\lambda_3$ represent suitable complex linear combinations of $\theta$'s
\begin{equation}
\lambda _{1_{n}}=\theta _{1_{n}}-i\theta _{2_{n}},\quad \lambda
_{2_{n}}=-\lambda _{1_{n}}^{*},\quad \lambda _{3_{n}}=\theta _{3_{n}}.
\label{lambdas}
\end{equation}
in order to guarantee that the transformation is unitary. The operator (\ref{gentransf}) generates the following unitary transformation:
\begin{eqnarray}
A_{n}^{\mu }\left( \theta \right)  &=&e^{-iG_{n}}a_{n}^{\mu
}e^{iG_{n}},\qquad \tilde{A}_{n}^{\mu }\left( \theta \right) = e^{-iG_{n}}\tilde{a}_{n}^{\mu
}e^{iG_{n}}\:,\nonumber \\
|0(\theta), p_v \rangle &=& e^{-iG}|0,p_v\rangle  \label{transf}
\end{eqnarray} 
and the structure of the Bogoliubov generator is similar to the operator used in the SU(1,1) formulation of the Thermo Field Dynamics \cite{gadelha,gadelha2,ume}. Writing the transformation for the creation/annihilation operators in the matrix notation, we have  

\begin{equation}
\left( 
\begin{array}{c}
A_{n}^{\mu }\left( \theta \right)  \\ 
\tilde{A}_{n}^{\mu \dagger }\left( \theta \right) 
\end{array}
\right) ={\mathcal B}_{n}\left( 
\begin{array}{c}
a_{n}^{\mu } \\ 
\tilde{a}_{n}^{\mu \dagger }
\end{array}
\right) ,
\label{doublet}
\end{equation}
where the explicit form of ${\mathcal B}_n$ operators is given by the following 
relation
\begin{equation}
{\mathcal B}_{n}=\cosh \left( i\Lambda _{n}\right) \mathbb{I} +\frac{\sinh \left(
i\Lambda _{n}\right) }{\left( i\Lambda _{n}\right) }\left( 
\begin{array}{cc}
i\lambda _{3_{n}} & i\lambda _{1_{n}} \\ 
i\lambda _{2_{n}} & -i\lambda _{3_{n}},
\end{array}
\right) 
\label{explB}
\end{equation}
and
\begin{equation}
\Lambda _{k}^{2}\equiv \left( \lambda _{3_{k}}^{2}+\lambda _{1_{k}}\lambda
_{2_{k}}\right) .
\label{biglambda}
\end{equation}

\noindent Now we just need to fix the parameters. If we choose the parameters such that
 
\begin{eqnarray}
 \cosh \left( i\Lambda_n \right) + 
 \frac{\lambda_{3_{n}}}{\Lambda_n }\mbox{sinh}\left( i\Lambda_n \right)&=&  f_n(t)\nonumber \\
\frac{\lambda _{1_{n}}}{\Lambda }\mbox{sinh}\left( i\Lambda_n \right) &=& g_n(t) \label{identif}
\end{eqnarray}
the transformation (\ref{doublet}) is exactly the same as (\ref{tbti}).

Let us now show the form of the Bogoliubov transformed vacuum. By applying the disentanglement theorem for $SU \left( 1,1 \right)$ \cite{cha,eber}, one can write the vacuum $|0(\theta), p_v \rangle$ in the following form
\begin{eqnarray}
 \left| 0\left( \theta \right) \right\rangle
=\prod_{n}e^{\Gamma _{1_{n}}\left( \tilde{a}_{n}^{\dagger }\cdot
a_{n}^{\dagger }\right) }e^{\log \left( \Gamma _{3_{n}}\right) \left(
a_{n}^{\dagger }\cdot a_{n}+\tilde{a}_{n}^{\dagger }\cdot \tilde{a}
_{n}+D\right) }e^{\Gamma _{2_{n}}
\left( a_{n}\cdot \tilde{a}_{n}\right)}\left| 0,p_v\right\rangle   \label{vaco}
\end{eqnarray}
where $D = tr(\delta_{ij})$ and
\begin{equation}
\Gamma _{1_{n}}=\frac{-\lambda _{1_{n}}\sinh \left( i\Lambda _{n}\right) }{
\Lambda _{n}\cosh \left( i\Lambda _{n}\right) +\lambda _{3_{n}}\sinh \left(
i\Lambda _{n}\right) },\quad \Gamma _{2_{n}}=\frac{\lambda _{2_{n}}\sinh
\left( i\Lambda _{n}\right) }{\Lambda _{n}\cosh \left( i\Lambda _{n}\right)
+\lambda _{3_{n}}\sinh \left( i\Lambda _{n}\right) },
\label{lambda12}
\end{equation}
\begin{equation}
\Gamma _{3_{n}}=\frac{\Lambda _{n}}{\Lambda _{n}\cosh \left( i\Lambda
_{n}\right) +\lambda _{3_{n}}\sinh \left( i\Lambda _{n}\right) },
\label{lambda3}
\end{equation}
 
\noindent Since the asymptotically flat vacuum is annihilated by $a_{n}^{i }$
and $\tilde{a}_{n}^{i}$, the equation (\ref{vaco}) can be rewritten as 
\begin{equation}
\left| 0(\theta ),p_v\right\rangle 
=\prod_{n}(\Gamma _{3_{n}})^{D}
e^{\Gamma _{1_{n}}\left( \tilde{a}_{n}^{\dagger
}\cdot a_{n}^{\dagger }\right) } \left| 0,p_v\right\rangle .
\label{thermvacfin}
\end{equation}
 
\noindent  Using the relations (\ref{identif}) one can see that
 \begin{equation}
 \Gamma _{1_{n}}= \frac{g^{\ast}_n (t)}{f_n (t)}
 \end{equation}
and we can write the vacuum $|0(\theta), p_v \rangle$ as the finite time vacuum  $|0(t), p_v \rangle$, which satisfies
\begin{equation}
{\cal A}^{i}_{n}(t)|0(t),p_v \rangle=
\tilde{\cal A}^{i}_{n}(t)|0(t), p_v \rangle =0.
\end{equation}

The Bogoliubov transformation maps the asymptotically flat string Hilbert space to its corresponding one at a finite $t$. In fact, the Bogoliubov operator is used to construct another representation of the Poisson algebra defined in (\ref{poisson}). This transformation can be interpreted as a relation between asymptotically flat observers and observers at a finite time. The late ones go with the string towards to the singularity. From the point of view of the asymptotically flat observers, the vacuum $|0(t),p_v\rangle$ is a superposition of $SU(1,1)$ coherent states, which has a particular structure of excitation, characterized by a condensation of right and left moving string modes. 

Now, let's see what happens with the state $|0(t),p_v\rangle $ at the singularity. It was shown in \cite{tsey} that it is possible to avoid the singularity using an analytical continuation of the Bessel functions, although this procedure leads to a discontinuity in the time derivative in the
zero mode sector \cite{tsey}. However, an important question was pointed out in \cite{gdn}. If we calculate the following projection
\begin{equation}
\langle p_{v}, 0 \mid 0(t),p_v \rangle= Ae^{-D
\sum_{n}\ln\left(1+|g_{n}(t)|^{2}\right)}
\label{ket}
\end{equation}

\noindent where $A$ is just a phase, and take the limit $t\rightarrow 0$ (which implies $n|t| \ll 1$), we get \linebreak
$\langle p_{v}, 0 \mid 0(t),p_v \rangle = 0$, showing that the state at the singularity ($t=0$) is unitarily inequivalent to the vacuum at $t = -\infty$. In other words, this result shows that close to the singularity the system defined by the 2d worldsheet quantum field theory is led to another representation of the canonical commutation relations, which is unitarily inequivalent to the representation at $t =-\infty$ \cite{ume}. The reason for the inequivalence between the Fock spaces is the condensation of an infinite number of string modes, including high frequency modes, in the  $\mid 0,p_v \rangle $ vacuum, when the string approaches the singularity. In the next section we are going to show that this result has a very suggestive interpretation: the unitarily inequivalence will be related to the appearance of different degrees of freedom - D-branes described in the weak coupling regime.

\section{The observer-dependent D-brane}
 
 Now we are going to show that, from the point of view of asymptotically flat observers, near to the singularity at $t=0$ (or in the limit $nt\ll 1$), we will have boundary state equations representing a ``transverse'' filling D-brane at this point. 

From equation  (\ref{osc}) and (\ref{fg}), the vacuum $|0(t), p_v \rangle$ is defined by

\begin{eqnarray}
{\cal A}^{i}_{n}(t)|0(t),p_v \rangle= \left(a^i_n + \frac{g^{\ast}_n (t)}{f_n (t)}\tilde{a}^{i\dagger}_n\right)|0(t),p_v \rangle = 0 \:,\nonumber\\
\tilde{\cal A}^{i}_{n}(t)|0(t), p_v \rangle = \left(\tilde{a}^i_n + \frac{g^{\ast}_n (t)}{f_n (t)} a^{i\dagger}_n\right)|0(t),p_v \rangle = 0. \label{vacuum}
\end{eqnarray}

\noindent In the limit $nt\ll 1$, $Z(2nt)$ goes to

\begin{equation}
Z(2nt) \approx -ie^{-i\frac{\pi}{2}\nu}\frac{\sqrt{\pi}(nt)^{1-\nu}}{\cos(\pi\nu)\Gamma(\frac{3}{2}-\nu)},\label{assint} 
\end{equation}
\noindent and equation (\ref{vacuum}) becomes
\begin{eqnarray}
\left(a^i_n + e^{i\pi\nu}\tilde{a}^{i\dagger}_n\right)|0(t=0) \rangle &= &0 \:, \nonumber\\
\left(\tilde{a}^i_n + e^{i\pi\nu} a^{i\dagger}_n\right)|0(t=0) \rangle &= &0 \:, \label{bondD}
\end{eqnarray}

\noindent where $|0(t=0) \rangle = \displaystyle\lim_{t\rightarrow0}|0(t), p_v \rangle $. Notice that the boundary state defined by the ``transverse'' filling D-brane at the singularity is a solution of the following equations:
\begin{eqnarray}
\left.\frac{\partial X^i}{\partial t} |B \rangle \right|_{nt\ll 1}&=&0 \nonumber \\
\tilde{x}|B \rangle &=&0\nonumber \\
\tilde{p}|B \rangle &=&0
\end{eqnarray}
and using the equations of motion, the first line can be written as:

\begin{eqnarray}
\left(a^i_n - \frac{\dot Z(2nt)}{Z^*(2nt)}\tilde{a}^{i\dagger}_n\right)|0(t=0) \rangle &= &0 \:, \nonumber\\
\left(\tilde{a}^i_n - \frac{\dot Z(2nt)}{Z^*(2nt)} a^{i\dagger}_n\right)|0(t=0) \rangle &= &0 \:.
\end{eqnarray}

Now, using the expression (\ref{assint}) the equations above become exactly the equation (\ref{bondD}). So, the vacuum at the singularity is seen by flat observers as the following D-brane boundary state
\begin{equation}
|0(t=0) \rangle= |B \rangle= {\it N}e^{e^{i\pi\nu}\sum a^{\dagger}_n\cdot \tilde{a}^{\dagger}_n}|0, p_v \rangle,
\end{equation}
where {\it N} is the usual normalization factor, which has the delta function that localizes the D-brane at the origin. 

\section{Conclusions}

Given type IIB superstring in pp-wave time dependent background, we analyzed how the flat vacuum $|0,p_v \rangle$ at $t=-\infty$ (which is annihilated by the $a_{n}^{i}$, $\tilde{a}_{n}^{i}$ modes) evolves to $t =0$, where a null singularity is located. We construct a Bogoluibov unitary $SU(1,1)$ transformation, relating the Hilbert space in $t=-\infty$ to the Hilbert space at a finite time. %In \cite{tsey}, a canonical transformation was performed in order to diagonalize the Hamiltonian in a finite time, writing it in terms of the time dependent modes ${\cal A}_{n}(t)$ and $\tilde{{\cal A}}_{n}(t)$. %In fact, this diagonalization process can be understood as a unitary $SU(1,1)$ Bogoliubov transformation, relating the Hilbert space in $t=-\infty$ to the Hilbert space at a finite time \cite{gdn}. The vacuum at a finite time $|0(t), p_v \rangle$ was obtained searching for the state that is annihilated by ${\cal A}_{n}(t)$ and $\tilde{{\cal A}}_{n}(t)$.

Owing to the Bogoliubov generator, the finite time vacuum, as seen by a flat observer, is a state with a particular structure of excitation, characterized by a condensation of right and left moving string modes. Besides, it was shown that, close to the singularity, the finite time Hilbert space is unitarily inequivalent to the flat Hilbert space: the vacuum at a finite time can't be a superposition of states which belong to the flat Hilbert space. The origin of the inequivalence is the condensation of infinite string modes in the flat vacuum when the string approaches the singularity.

In the last section we interpreted this inequivalent vacuum. We show that, for asymptotically flat observers, the closed string vacuum close to the singularity appears as a boundary state, which is in fact a D-brane described in the closed string channel. On the other hand, for observers who go with the string towards to the singularity, the vacuum at $t=0$ is the original vacuum, which means that this result is observer dependent. This is a very suggestive result: as it was pointed out in \cite{tsey}, a D-brane at the singularity is necessary in order to the closed string passes through it without discontinuities in the zero mode. In addition, due to the fact that the pp-wave time dependent background comes from the Penrose limit of D-branes localized at the singularity in Brinkmann coordinates, it is natural to expect a D-brane at this point. Furthermore, although there is no horizon in this model, our result is closely related to the stretched horizon paradigm and complementary principle of black holes \cite{suss1}. The simple result presented here for free closed strings suggests the same behavior for strings close to the null cosmological singularity. Some related ideas were pointed out in \cite{sumit}.

Finally, it should be emphasized that this is a toy model to show how interesting string theory ideas can be implemented in the weak coupling regime.  A more realistic and involved situation should take into account non-perturbative results.

\acknowledgments The authors would like to thank Alexandre Leite Gadelha and Marcelo Botta for useful discussions.


\begin{thebibliography}{99}
\bibitem {suss1} L. Susskind, L. Thorlacius and J. Uglum, ``The Stretched Horizon And Black Hole Complementarity'', Phys. Rev. D {\bf 48} (1993) 3743 [arXiv:hep-th/9306069]; L. Susskind, ``Strings, black holes and Lorentz contraction'', Phys. Rev. D {\bf 49} (1994) 6606 [arXiv:hep-th/9308139]; L. Susskind, ``Some speculations about black hole entropy in string theory'', arXiv:hep-th/9309145.

  
\bibitem{tsey} G. Papadopoulos, J. G. Russo and A. A. Tseytlin, ``Solvable model of strings in a time-dependent plane-wave background'', Class. Quant. Grav. {\bf 20} (2003) 969 [arXiv:hep-th/0211289].

\bibitem{Duval} C. Duval, G. W. Gibbons and P. Horvathy, ``Celestial Mechanics, Conformal Structures, and Gravitational Waves'', Phys. Rev. D {\bf 43} (1991) 3907 [arXiv:hep-th/0512188].

\bibitem{amati} D. Amati and C. Klimcik, ``Strings in a shock
wave background and generation of curved geometry from flat
space string theory'', Phys. Lett. B {\bf 210} (1988) 92.

\bibitem{horo} G. T. Horowitz and A. R. Steif, ``Strings in
strong gravitational fields'', Phys. Rev. D {\bf 42} (1990) 1950;
``Space-time singularities in string theory'', Phys. Rev. Lett.
{\bf 64} (1990) 260.

\bibitem{brooks} R. Brooks, ``Plane wave gravitons, curvature
singularities and string physics'', Mod. Phys. Lett. A {\bf 6} (1991) 841.

\bibitem{sanchez} H. J. de Vega, M. Ramon Medrano and N. Sanchez,
``Classical and quantum strings near space-time singularities:
gravitational plane waves with arbitrary polarization'', Class.
Quant. Grav. {\bf 10} (1993) 2007.

\bibitem{gaspvene} M. Gasperini and G. Veneziano, ``The pre-big bang scenario in string cosmology'', Phys. Rept. {\bf 373} (2003) 1 [arXiv:hep-th/0207130].


\bibitem{Borunda} M. Blau, M. Borunda, M. O'Loughlin and G. Papadopoulos, ``Penrose limits and spacetime singularities'', Class. Quant. Grav. {\bf 21} (2004) L43 [arXiv:hep-th/0312029]; M. Blau, M. Borunda, M. O'Loughlin and G. Papadopoulos, ``The universality of Penrose limits near space-time singularities'', JHEP {\bf 0407} (2004) 068 [arXiv:hep-th/0403252].


\bibitem{gdn} A. L. Gadelha, D. Z. Marchioro and D. L. Nedel, ``Entanglement and entropy operator for strings in pp-wave time dependent background'', Phys. Lett. B {\bf 639} (2006) 383 [arXiv:hep-th/0605237].


\bibitem{gadelha} M. C. B. Abdalla and A. L. Gadelha, ``General unitary SU(1,1) TFD formulation'', Phys. Lett. A {\bf 322} (2004) 31 [arXiv:hep-th/0309254]; M. C. B. Abdalla, A. L. Gadelha and D. L. Nedel, ``General unitary TFD formulation for superstrings'', PoS {\bf WC2004} (2004) 032 [arXiv:hep-th/0412128].


\bibitem{gadelha2} M. C. B. Abdalla, A. L. Gadelha and I. V. Vancea, ``On the SU(1,1) thermal group of bosonic strings and D-branes'', Phys. Rev. D {\bf 66} (2002) 065005 [arXiv:hep-th/0203222].

\bibitem{ume} H. Umezawa, {\em Advanced Field Theory: Micro, Macro and Thermal Field}, American Institute of Physics Press 1995.

\bibitem{sumit} S. R. Das and J. Michelson, ``Matrix membrane big bangs and D-brane production'', Phys. Rev. D {\bf 73} (2006) 126006 [arXiv:hep-th/0602099].



\bibitem{polch} P. Di Vecchia, A. Liccardo, ``D branes in string theory. I'', NATO Adv. Study Inst. Ser. C. Math. Phys. Sci. {\bf 556} (2000) 1  [arXiv:hep-th/9912161].


\bibitem{cha} S. Chaturvedi and V. Srinivasan, J. Phys. A {\bf 32} (1999) 1909.

\bibitem{eber} K. W\'{o}dkiewicz and J. H. Eberly, J. Opt. Soc. Am. B {\bf 3} (1985) 485.

\end{thebibliography}
\end{document}